\documentclass[final,1p,times]{elsarticle}

\newcommand{\degC}{$^{\circ}$C~}

\newcommand{\neut}{$\rm n_{eq}/cm^2$}

\usepackage{amssymb}
\usepackage{amsmath}
\usepackage{lineno}
\usepackage[hidelinks]{hyperref}

\journal{Nuclear Instruments and Methods in Physics Research Section A}

\begin{document}

\begin{frontmatter}



\title{The CMS Barrel Timing Layer: test beam confirmation of module timing performance}


%\author{Full BTL authorlist under construction} %% Author name
 \author[21]{F.~Addesa \fnref{francesca}}
 \author[23,24]{P.~Akrap}
 \author[1]{A.~Albert}
 \author[6]{B.~Allmond}
 \author[32]{T.~Anderson}
 \author[28,29]{J.~Babbar}
 \author[3]{D.~Baranyai}
 \author[23]{P.~Barria \fnref{patrizia}}
 \author[23,24]{C.~Basile}
 \author[10]{A.~Benaglia}

% ...
\affiliation[1]{organization = {California Institute of Technology},
               addressline = {1200 East California Boulevard},
               city = { Pasadena California 91125},
               country = { United States of America}}

% ...
\affiliation[3]{organization = {University of Debrecen, Faculty of Informatics},
               addressline = {Kassai str 26},
               city = { Debrecen 4032},
               country = { Hungary}}

% ...
\affiliation[6]{organization = {Kansas State University},
               addressline = {705 N Martin Luther King Jr Drive},
               city = { Manhattan Kansas 66502},
               country = { United States of America}}

% ...
\affiliation[10]{organization = {INFN Sezione di Milano-Bicocca},
               addressline = {Piazza della Scienza 3},
               city = { 20126 Milano},
               country = { Italy}}

% ...
\affiliation[21]{organization = {Princeton University},
               addressline = {Princeton},
               city = { New Jersey 08544},
               country = { United States of America}}

% ...
\affiliation[23]{organization = {INFN Sezione di Roma},
               addressline = {Piazzale A. Moro 2},
               city = { 00185 Roma},
               country = { Italy}}

 \affiliation[24]{organization = {Sapienza Universita' di Roma},
               addressline = {Piazzale A. Moro 2},
               city = { 00185 Roma},
               country = { Italy}}

% ...
\affiliation[28]{organization = {INFN Sezione di Trieste},
               addressline = {via A. Valerio 2},
               city = { 34127 Trieste},
               country = { Italy}}

 \affiliation[29]{organization = {Università di Trieste},
               addressline = {via A. Valerio 2},
               city = { 34127 Trieste},
               country = { Italy}}

% ...
\affiliation[32]{organization = {University of Virginia },
               addressline = {Charlottesville},
               city = { Virginia},
               country = { United States of America}}

\fntext[francesca]{Now at Paul Scherrer Institute PSI, Villigen, Switzerland}
\fntext[patrizia]{Now at IAPS - INAF, Rome, Italy}
\fntext[yuri]{Also at Institute for Nuclear Research, Moscow, Russia}
\fntext[badder]{Now at University of Minnesota, Minneapolis, USA}
\fntext[paolo]{Now at INFN Sezione di Torino and Università degli Studi di Torino, Torino, Italy}
\fntext[samuele]{Now at Université Paris-Saclay, Gif-sur-Yvette, France}
\fntext[licheng]{Now at University of Maryland, Maryland, USA}

%\affiliation{organization={},%Department and Organization

%% Abstract
\begin{abstract}
First of its kind, the barrel section of the MIP Timing Detector is a large area timing detector based on LYSO:Ce crystals and SiPMs which are required to operate in an unprecedentedly harsh radiation environment (up to an integrated fluence of $2\times10^{14}$~1~MeV~\neut). 
It is designed as a key element of the upgrade of the existing CMS detector to provide a time resolution for minimum ionizing particles in the range between $30-60$~ps throughout the entire operation at the High Luminosity LHC.
A thorough optimization of its components has led to the final detector module layout which exploits 25~$\rm \mu m$ cell size SiPMs and 3.75~mm thick crystals. This design achieved the target performance in a series of test beam campaigns. 
In this paper we present test beam results which demonstrate the desired performance of detector modules in terms of radiation tolerance, time resolution and response uniformity.
\end{abstract}

%% \begin{graphicalabstract}
%% \includegraphics{grabs}
%% \end{graphicalabstract}

%% \begin{highlights}
%% \item Research highlight 1
%% \item Research highlight 2
%% \end{highlights}

%% Keywords

\end{frontmatter}



\section{Introduction}
\label{sec:introduction}
%% Labels are used to cross-reference an item using \ref command.
The MIP Timing Detector (MTD)~\cite{CMS_MTD_TDR} of the CMS experiment~\cite{CMS:2008xjf} is designed to measure the time of arrival of minimum ionizing particles (MIPs) with a resolution ranging from about 30~ps, at the beginning of the high luminosity phase of the LHC (HL-LHC), to approximately 60~ps in the barrel part, by the end of the detector operation. 
This level of precision in time-tagging charged particles from collision events will significantly enhance CMS performance in the challenging conditions of the HL-LHC.
A timing resolution for charged particles significantly smaller than the temporal spread of the luminous region (approximately 200~ps RMS), will help separate multiple interactions that occur in the same bunch crossing. This will improve pileup rejection and effectively recover event reconstruction quality to the level achieved in the current LHC. Additionally, time-of-flight information will provide new capabilities to CMS, including particle identification of low momentum charged hadrons and extending the potential of searches for long-lived particles~\cite{CMS_MTD_TDR}~\cite{CMS-DP-2022-025}.

The structure of the barrel timing layer (BTL) is described in detail in the Technical Design Report (TDR) \cite{CMS_MTD_TDR}. The BTL consists of a cylindrical layer of 5200~mm length and approximately 1150~mm radius, placed between the CMS tracker and the electromagnetic calorimeter and covering a surface of about 38~m$^2$. The active element is the \emph{sensor module}, an array of 16 LYSO:Ce crystal bars coupled to silicon photomultipliers (SiPMs). The BTL will consist of 10368 sensor modules, for a total of 331776 readout channels, two per crystal, and will cover the pseudorapidity region up to $|\eta| <$ 1.48.

Since the TDR, substantial R\&D on various detector components has occurred to optimize the BTL layout including the development of the final version of the readout ASIC (TOFHIR2)~\cite{albuquerque2024tofhir2}, the integration of thermoelectric coolers (TECs) on the SiPM array package to operate the SiPMs at a temperature of T$_{op} = -45$\degC and perform in-situ annealing at T$_{a} = 60$\degC during the HL-LHC stops \cite{Bornheim_2023}.
An extensive optimization of detector sensor modules (crystals and SiPMs) has also been conducted through a series of test beam campaigns~\cite{Addesa:2024mpu}, leading to the identification of the final sensor module specifications.

In this paper, we present a detailed characterization of the response of BTL final sensor module prototypes to minimum ionizing particles performed using 180~GeV pions from the CERN SPS beam line.

The time resolution of both non-irradiated modules and ones irradiated up to the integrated fluence expected at the end of operation is presented, as well as the uniformity of the module response. The results demonstrate that the designed time resolution can be achieved over the entire detector lifetime during HL-LHC operation: specifically remaining better than 60 ps (up to a '1 MeV neutron equivalent fluence', \neut, of $2\times 10^{14}$\neut), while operating within the available power budget of 50~mW per SiPM \cite{Bornheim_2023}.

\section{Description of the prototypes}\label{sec:modules} 
\begin{figure}[!tbp]
    \centering
    \includegraphics[width=0.99\linewidth]{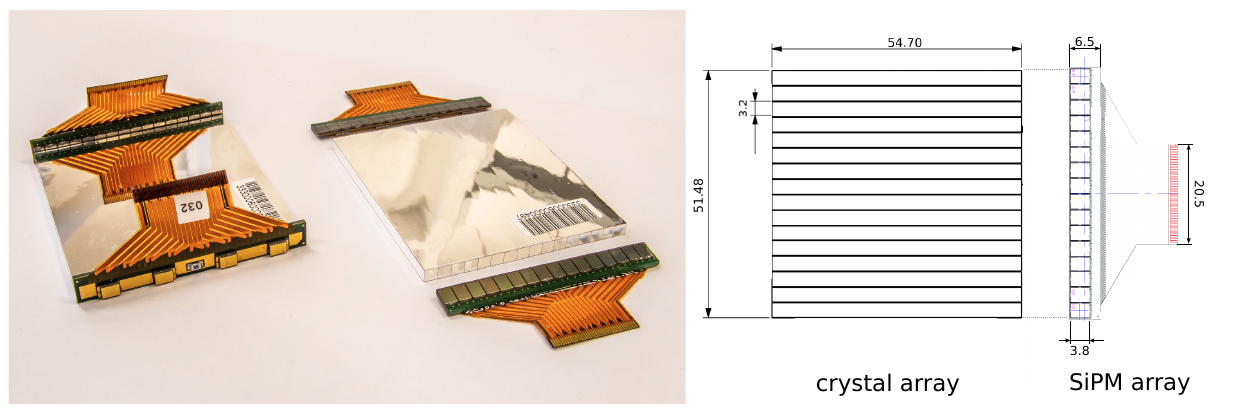}        
    \caption{Picture of a BTL sensor module after and before gluing between crystals and SiPMs (left) with dimensions of the crystal and SiPM arrays (right). Units of dimensions are in mm.}
\label{fig:sm_pic}
\end{figure}

A BTL sensor module, illustrated in figure~\ref{fig:sm_pic}, consists of 16 LYSO:Ce crystal bars, each with dimensions of $\rm 54.7 \times3.12 \times 3.75~mm^3$, and covered on lateral faces by a 80~$\mathrm{\mu}$m thick reflector such that the pitch between two adjacent crystals is 3.2~mm. Each end of the module is read out by an array of 16 SiPMs. The active area of each SiPM is $2.91\times3.80$~mm and is aligned with the center of the crystal.
The crystal arrays used in these studies were manufactured by Sichuan Tianle Photonics (STP) and Suzhou JT Crystal Technology (JTC).
These crystals feature an average decay time of 43~ns. %The planarity of the crystal faces is better than 20~$\mathrm{\mu}$m. 
More details on the crystal characterization are provided in \cite{Addesa:2024mpu}.

Each bar is wrapped in a thin layer of Enhanced Specular Reflector (3M ESR) which provides isolation of each channel minimizing optical cross talk between adjacent crystals.
The SiPM arrays are coupled to the LYSO:Ce crystals by means of a 100~$\mathrm{\mu}$m layer of RTV3145 glue. Each SiPM array package includes a PT1000 temperature sensor and four TECs~\cite{Bornheim_2023} for temperature control and stabilization.

The SiPMs used for BTL have dimensions matching the crystal end-face for optimal light collection, a cell-size of 25~$\mathrm{\mu}$m and are manufactured by Hamamatsu Photonics (HPK). They feature a Photon Detection Efficiency (PDE) of 57 (30)\% and a gain of 10$^6$ ($3.6 \times 10^5$) for over-voltage \footnote{Excess bias beyond the break-down voltage} V$_{OV}$~=~3.5 (1)~V, typical of the BTL beginning (end) of operation. In the sensor module, the light output reaches approximately 2400 photoelectrons (pe) per MeV when SiPMs are operated at $V_{OV}$~=~3.5~V.

The choice of 25~$\mathrm{\mu}$m as optimal cell-size was based on a trade-off between two effects: PDE and Dark Count Rate (DCR). Larger cell sizes result in larger PDE and gain thus providing larger signals which help reducing the photo-statistics and electronics noise contributions to the time resolution. On the other end, larger cell-size also implies larger DCR and power dissipation for irradiated SiPMs.
A detailed study of the dependencies of the time resolution on the crystal and SiPM parameters, which informed the choice of the final BTL sensors layout, is reported in~\cite{Addesa:2024mpu}.

Four module prototypes, one with non-irradiated SiPMs and three with SiPMs irradiated to different fluences, were tested to evaluate the BTL performance at different stages over the BTL lifetime under expected HL-LHC conditions.
The irradiation of the SiPM arrays was performed at the JSI neutron reactor in Ljubjana for integrated fluences of 1~$\times$~10$^{13}$, 1~$\times$~10$^{14}$, 2~$\times$~10$^{14}$ \neut, corresponding to the radiation levels expected after about 150, 1500 and 3000~fb$^{-1}$, respectively.
The uncertainty on the irradiation levels is estimated to be approximately 10\% based on the comparison of several SiPMs exposed to the same nominal fluence during different irradiation campaigns at the Ljubljana reactor. 
The combination of SiPM annealing history and operation temperature at the test beam (T$_{TB}$ = -35\degC) was chosen to reproduce the same level of DCR expected for the BTL detector after the same level of irradiation with in situ operation at T$_{op}$ = -45\degC and SiPM annealing at T$_{a} = 60$\degC. All SiPMs were annealed for 40 minutes at 70\degC, three days at 110\degC and four days at 120\degC to reproduce, within a 10\% uncertainty, the level of thermal annealing expected in the BTL detector during its operation~\cite{Addesa:2024mpu}. 

\section{Experimental setup and procedures}
\label{sec:experimentalSetup}

A test beam campaign was conducted at the CERN SPS H8 beam line, using 180~GeV pions, to evaluate the time resolution of the sensor modules in MIP detection. The experimental setup and procedures used in this study are the same as those described in a previous work~\cite{Addesa:2024mpu}.

Sensor modules were tested with front-end test boards using TOFHIR2 ASICs~\cite{albuquerque2024tofhir2} and read out via a FEB/D board. TOFHIR2 provides measurements of the time of arrival of the MIP signals, using 20~ps TDC \footnote{Time to Digital Converter} binning and a leading-edge current discriminator with configurable threshold, and of the amplitude of the signals through charge integration.

\begin{figure}[!tbp]
    \centering
    \includegraphics[width=0.99\linewidth]{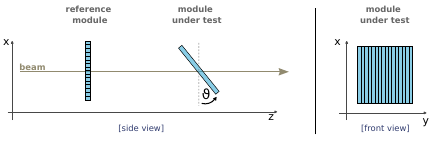}
    \caption{Schematic representation of the sensor module orientations along the beam line: side view (left) and front view (right).}
\label{fig:setup_sketch}
\end{figure}

A reference (non-irradiated) sensor module was positioned at normal incidence to the beam to provide coarse position data, while the device under test (DUT) was tilted by an angle $\theta$ using a remotely controlled stage to simulate energy deposits expected in operational conditions, as shown in the sketch in figure~\ref{fig:setup_sketch}. 
The setup temperature was stabilized at -30\degC within a light-tight cold box, while TECs were used to adjust the SiPM temperature to T$_{TB}$ = -35\degC. The SiPM temperature, measured using the PT1000 sensors on the SiPM arrays, remained stable to within 1\degC throughout the entire test beam.

Events are selected based on energy deposition in a single crystal bar, consistent with MIP behavior. The energy, averaged between the measurements from the two SiPMs located on the ends of each bar, is used to select events within a window around the most probable value (MPV) of the MIP energy distribution. An additional selection on the impact point is applied by requiring a MIP event in a central bar of the upstream reference module, reducing the beam spot size to a few millimeters along the bar's longitudinal axis.

In BTL, the time of arrival of a MIP in each bar is calculated as the average of the time measurements at the two ends. In this work, the bar time resolution is estimated by following the approach described in~\cite{Addesa:2024mpu}, which involves taking half the spread of the time difference between the signals from each end. For a fixed impact point position along the bar and assuming no correlated uncertainties between the two time measurements, this quantity is equivalent to the resolution of the average time. 
It was also confirmed in ~\cite{Addesa:2024mpu} that a direct measurement of the bar resolution relative to a high quality reference yields the same result.
The measured time resolution is affected by the residual dependence of the time difference on the impact point, which in turn depends on the resolution of the impact point. This resolution is determined by the selected bar in the reference module and therefore varies with the tilt angle ($\theta$). This contribution is estimated to be 10 ps for $\theta=32^{o}$, 13 ps for $\theta=52^{o}$, and 18 ps for $\theta=64^{o}$.
A scan of the leading edge discrimination threshold is conducted to determine the optimal value, i.e. the one providing the best time resolution, for each sensor configuration.

\section{Results}\label{sec:results}
\begin{figure}[!h]
    \centering
    \includegraphics[width=0.49\linewidth]{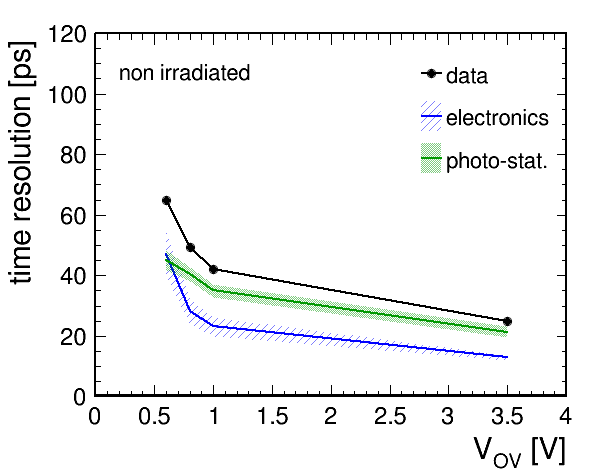}
    \includegraphics[width=0.49\linewidth]{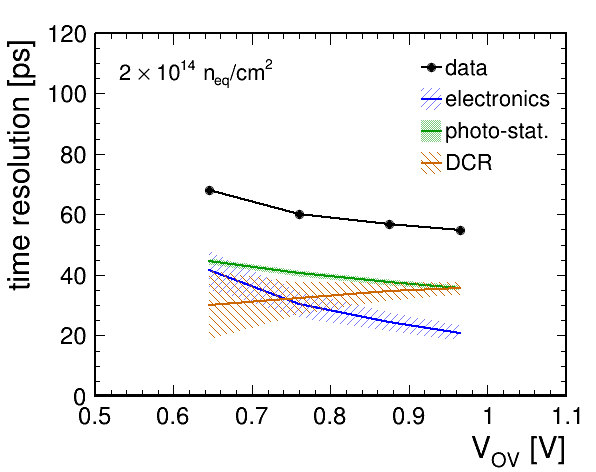}

    \caption{Time resolution as a function of the SiPM V$_{OV}$ for modules with non-irradiated SiPMs (left) and SiPMs irradiated to 2~$\times$~10$^{14}$~\neut~(right). The time resolution measured with beam data is shown with black dots, the main individual contributions to the time resolution are shown by the colored lines: electronics (blue), stochastic (green), and DCR (orange). The DCR contribution is completely negligible for non-irradiated SiPMs.}
    \label{fig:timeResolution_components}

\end{figure}

The time resolution as a function of the SiPM over-voltage is reported in figure~\ref{fig:timeResolution_components} for one module with non-irradiated SiPMs (left) and for another one with SiPMs irradiated to 2$\times$10$^{14}$ \neut (right). The modules under test were tilted by $\theta$~=~52$^\circ$ relative to the beam to reproduce the MPV of the energy deposit expected in the central part of the BTL, corresponding to 5.2 MeV, from tracks in collision events. In this configuration, the module is oriented such that a MIP crosses a single crystal (see figure~\ref{fig:setup_sketch}).

The main individual contributions to the measured time resolution, due to the electronic noise, photo-statistics and DCR noise, are also shown in figure~\ref{fig:timeResolution_components}, and were estimated according to the procedure described in~\cite{Addesa:2024mpu}. The electronics noise contribution, which depends on the slope of the pulse at the discriminator threshold, becomes significant at low over-voltage values due to the decrease in SiPM gain and PDE with lower over-voltages. As the over-voltage increases, the dominant contribution is the photo-statistic one, while for irradiated SiPMs, the DCR plays a significant role. The DCR not only limits the time resolution but also impacts operation due to the large power dissipation and self-heating of the SiPMs.
A time resolution of 25 ps is achieved at an over-voltage of approximately 3.5 V for non-irradiated SiPMs, representative of the beginning of BTL operation, while a resolution of 55 ps is obtained at 1 V for irradiated SiPMs, corresponding to the end of operation conditions.
The DCR per SiPM for the irradiated case at V$_{OV}$~=~1~V is about 20~GHz. The operation of the SiPMs above this voltage is limited by power budget constraints and self-heating effects.

\begin{figure}[!tbp]
    \includegraphics[width=0.495\linewidth]{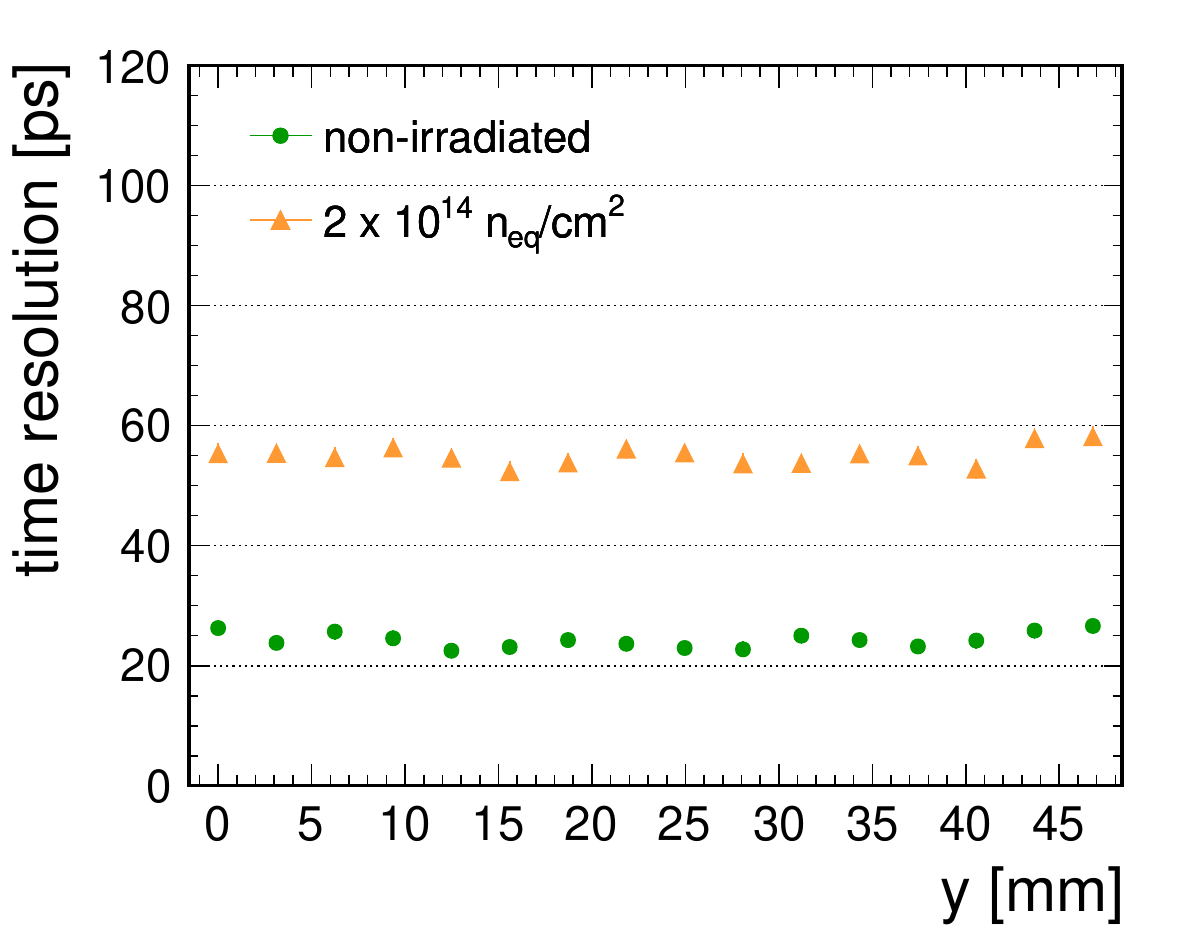}
    \includegraphics[width=0.495\linewidth]{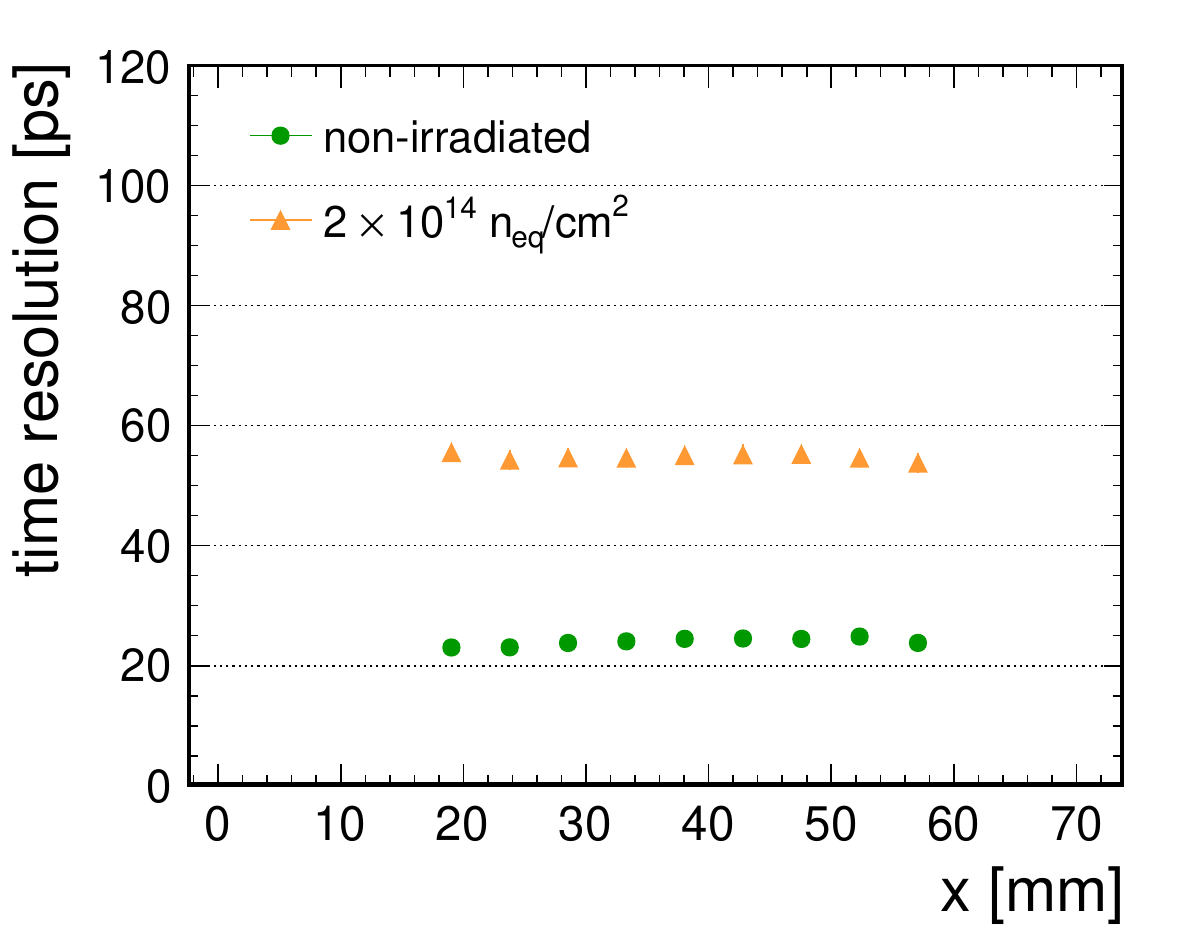}
    \caption{On the left, internal time resolution uniformity for one module with non-irradiated SiPMs operated at V$_{OV}$~=~3.5~V  and another with SiPMs irradiated to a fluence of 2~$\times$~10$^{14}$~\neut~ operated at V$_{OV}$~=~0.96~V.
    X and y are defined according to figure~\ref{fig:setup_sketch}.
    On the right, time resolution uniformity along the bar longitudinal axis ($x$) of the same two modules. The time resolution for each impact point is averaged over all the bars of a module. Due to the coarse determination of the x position, measurements do not  cover the first and last 5 mm of the bar.
    }
    \label{fig:module_uniformity}
\end{figure}

A sensor module covers a surface of about $52\times55$~mm$^2$ as shown in figure~\ref{fig:sm_pic}. The response uniformity within a module was quantitatively assessed by evaluating the spread of the time resolution measured on the 16 crystal bars and for different MIP impact point positions along the longitudinal axis of the bars.
Results are compared in figure~\ref{fig:module_uniformity} for a module with non-irradiated SiPMs and a module with SiPM arrays irradiated to a fluence of 2$\times$10$^{14}$~\neut. 
The spread of time resolution across different bars is less than 2~ps RMS for both the non-irradiated and irradiated modules. This indicates a uniform light output across the bars in the module, effective optical coupling of all SiPMs to the crystal bars, and confirms that the variation in SiPM breakdown voltages within a single array (nominally within 150 mV) does not affect the uniformity of time resolution across the module.
Along the $x$ direction, i.e. the longitudinal axis of a crystal bar, uniformity was studied by selecting events in which a MIP interacted in different bars of the upstream reference module. These interactions correspond to steps of approximately 5~mm due to the tilt angle of the module under test (52$^\circ$ relative to the beam).
The uniformity of time resolution along the bar is also better than 2~ps for both non-irradiated and irradiated modules. 

In the BTL detector, sensor modules are positioned at various locations along the barrel axis, resulting in a variation in the mean angle at which particles impact the modules. Specifically, sensor modules at higher pseudorapidity ($\eta$) will be traversed by particles at larger angles, leading to higher energy deposition and, consequently, better time resolution. However, this benefit is partially offset by an increase in radiation levels, which rise by about 20\% along the length of the detector (from about $1.65\times10^{14}$ to $1.90\times10^{14}$ \neut ~\cite{CMS_MTD_TDR} when going from $|\eta|=0$ to $|\eta|=1.45$).

\begin{figure}[!tbp]
    \centering
    \includegraphics[width=0.49\linewidth]{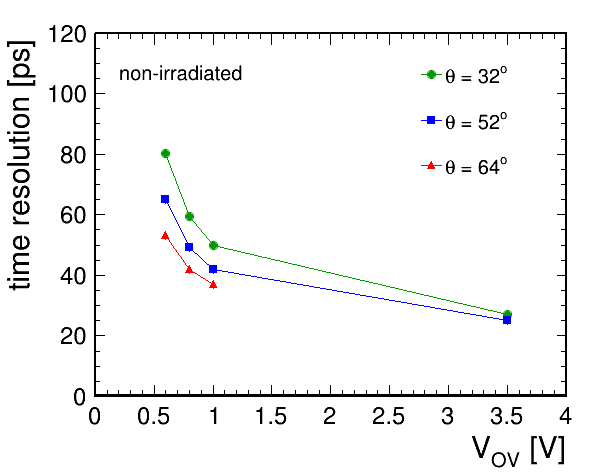}
    \includegraphics[width=0.49\linewidth]{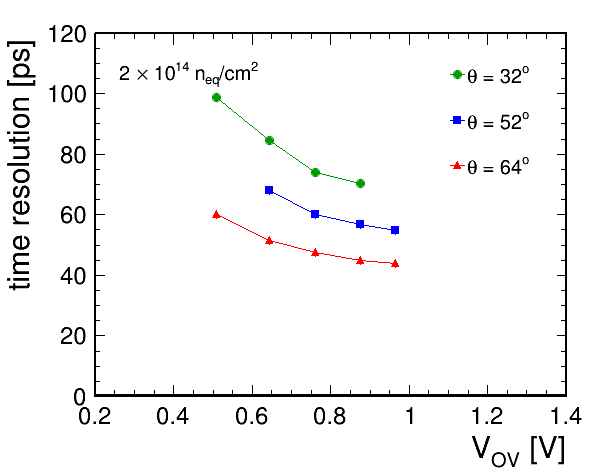}
    \caption{Time resolution as a function of the over-voltage for different impact angles to the beam direction: module with non-irradiated SiPMs (left) and with SiPMs irradiated to a fluence of 2$\times$10$^{14}$~\neut~ (right). The missing point at high over-voltage for non-irradiated SiPMs at $\theta=64^{o}$ is due to the poor data quality of the corresponding dataset.}
\label{fig:timeResolution_angles}
\end{figure}

To assess the impact of these factors on time resolution, we studied the performance of three sensor modules: one with non-irradiated SiPMs, one with SiPMs irradiated to $1 \times 10^{14}$~\neut~and one with SiPMs irradiated to $2 \times 10^{14}$~\neut, which covers the maximum irradiation level expected for SiPMs within BTL.
The time resolution was measured for different impact angles of the MIP, as shown in figure~\ref{fig:timeResolution_angles}, for both the non-irradiated module and for the module irradiated to $2 \times 10^{14}$~\neut. The tilt angles ($\theta=32^{\circ}$, 52$^{\circ}$, 64$^{\circ}$) were selected to represent the most probable MIP energy deposition values across the low, medium, and high pseudorapidity regions of the BTL. 
The time resolution of the modules is shown in figure~\ref{fig:timeResolution_angles} as a function of the SiPM over-voltage, $V_{OV}$, within the operational range compatible with power budget constraints. The results highlight how, at larger angles, the increase in energy deposition and thus in the number of photoelectrons, $N_{pe}$, impacts the time resolution. It can be noted that the relative gain in time resolution at higher pseudorapidity is more pronounced for irradiated SiPMs for which the DCR term, scaling as $1/N_{pe}$ becomes sizable.

\begin{figure}[!tbp]
    \centering
    \includegraphics[width=0.49\linewidth]{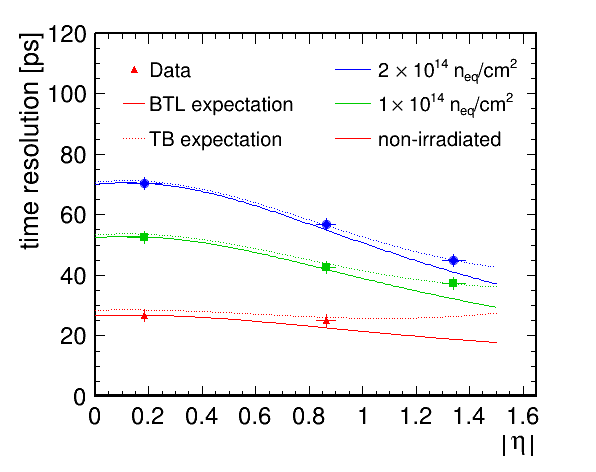}
    \includegraphics[width=0.49\linewidth]{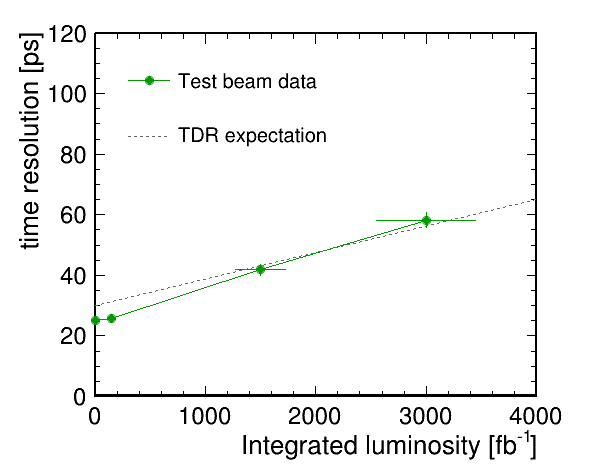}
    \caption{Left: Time resolution as a function of the equivalent pseudorapidity for modules irradiated to different fluences. The data (dots) are compared with the BTL model expectations (solid lines). The dashed lines correspond to the resolution expected for the test beam data (TB expectation), obtained by adding to the BTL model expectations the extra contribution due to the residual dependence of the time difference on the impact point described in section~\ref{sec:experimentalSetup}. Right: Time resolution as a function of the equivalent integrated luminosity. The data are from test beam measurements of modules with non-irradiated SiPMs and SiPM arrays irradiated to different fluences. The dotted line is the target time resolution from the TDR.  
    }
    \label{fig:timeResolution_lumi_eta_summary}
\end{figure}

The angles used in the test beam ($\theta)$ have been mapped into the corresponding locations across the BTL detector ($\eta$), that are equivalent in terms of energy deposited by a MIP in the crystal. The best time resolution achieved within the allowed power budget is then shown in the left panel of figure~\ref{fig:timeResolution_lumi_eta_summary} as a function of the pseudorapidity for both non-irradiated SiPMs and for SiPMs irradiated to $1\times 10^{14}$~\neut, $2\times 10^{14}$~\neut, which correspond to beginning, half and end of the detector operation, respectively.
The variation of time resolution as a function of pseudorapidity based on the scaling of various contributions to the time resolution on the number of photoelectrons according to the model described in~\cite{Addesa:2024mpu} is also shown (after normalization to experimental measurement at $\eta = 0.2$) as a continuous colored line, and compared with data. The small discrepancy at large pseudorapidity between the measured time resolution and that estimated from the model is ascribed to the additional contribution to the time resolution measured from test beam data, due to the residual time difference dependence on the impact point, as discussed in section~\ref{sec:experimentalSetup}. The agreement is recovered once this term is added in quadrature, as shown by the dashed line in figure~\ref{fig:timeResolution_lumi_eta_summary} (left).

The time resolution measured for each module, operating under the optimal (nominal) over-voltage, is shown in figure~\ref{fig:timeResolution_lumi_eta_summary} (right) as a function of the equivalent integrated luminosity. The uncertainty on the integrated luminosity is estimated to be about 15\% from the uncertainties in the irradiation and in the annealing model (see section~\ref{sec:modules}). The experimental results are compared with the target time resolution foreseen in the MTD TDR \cite{CMS_MTD_TDR}, where the impact of the MTD detector on the HL-LHC physics goals is also assessed.
These results demonstrate that the sensor design can maintain the desired performance throughout the entire operation of the HL-LHC. Despite the challenging radiation levels, the system is expected to maintain a time resolution below 60 ps at high fluence by the end of operation.

\section{Conclusion}
A set of BTL sensor modules, constructed of non-irradiated SiPMs and SiPMs irradiated to different levels of fluences (up to $2\times 10^{14}$~1~MeV~\neut), have been tested at the SPS CERN beam line with 180 GeV pions. The results collected have demonstrated that a time resolution of about 25~ps is achieved with non-irradiated modules operating at 3.5~V over-voltage. The time resolution degrades smoothly to about 55~ps for modules irradiated to the maximum fluence anticipated at the end of the detector operation, when the optimal SiPM over-voltage is about 1~V. The response of a sensor module was proven to be uniform to less than 2~ps RMS over its entire active surface of about $52\times55 $~mm$^2$ for both irradiated and non-irradiated modules. A study of the module performance as a function of the particle impact angle was also performed to vary the amount of energy deposited in the crystals and assess the corresponding variation in time resolution, emulating the performance of sensor modules that will be located at various pseudorapidity regions in the final detector.
Overall, the results presented herein prove that, with the final design and technical specifications, the BTL sensor modules can reach the target time resolution required for the detector to meet its design physics goals \cite{CMS_MTD_TDR} during the HL-LHC operation.

\vspace{1cm}

\textit{Acknowledgments: The authors are grateful to the technical experts of the CERN beam line facilities for their invaluable help.}\\

\textit{Funding: This project has received funding from the European Union’s Horizon Europe Research and Innovation programme under Grant Agreement No. 101057511 (EURO-LABS), the European Union - Next Generation EU  Mission 4 Component 2 CUP I53D23001520006, MoST (China) under the National Key R\&D Program of China (No. 2022YFA1602100), the NSFC (China) under global scientific research funding projects (No. W2443006), the Funda\c{c}\~ao para a Ci\^encia e a Tecnologia (FCT), Portugal and ICSC – National Research Center for High Performance Computing, Big Data and Quantum Computing funded by the NextGenerationEU program (Italy), Latvian Council of Sciences State research programme project VPP-IZM-CERN-2022/1-0001.}

\end{document}